\documentclass{aa}
\usepackage[utf8]{inputenc}
\usepackage{natbib}
\usepackage{graphicx}
\usepackage{fabien}
\usepackage{braket}

\usepackage[breaklinks,colorlinks,citecolor=blue]{hyperref}

\usepackage{natbib}
\bibpunct{(}{)}{;}{a}{}{,}
\usepackage{verbatim}

\begin{document}
        
        \title{Efficient computation of the super-sample covariance for \\ stage IV galaxy surveys}
        \titlerunning{Efficient SSC for stage IV surveys}
        
        \author{
                Fabien Lacasa\inst{\ref{Unipd},\ref{INFN-Padova},\ref{Unige},\ref{IAS}},
                Marie Aubert\inst{\ref{Lyon}}
                Philippe Baratta\inst{\ref{CPPM}},
                Julien Carron\inst{\ref{Unige}},
                Adélie Gorce\inst{\ref{McGill}}
                Sylvain Gouyou Beauchamps\inst{\ref{CPPM}},
                Louis Legrand\inst{\ref{Unige}},
                Azadeh Moradinezhad Dizgah\inst{\ref{Unige}},
                Isaac Tutusaus\inst{\ref{IRAP},\ref{Unige}}
        }
        \authorrunning{F. Lacasa et al.}
        
        \institute{
                Dipartimento di Fisica e Astronomia ``Galileo Galilei'', Università degli Studi di Padova, via Marzolo 8, I-35131, Padova, Italy\label{Unipd}
                \and INFN, Sezione di Padova, via Marzolo 8, I-35131, Padova, Italy\label{INFN-Padova}
                \and D\'{e}partement de Physique Th\'{e}orique and Center for Astroparticle Physics, Universit\'{e} de Gen\`{e}ve, 24 quai Ernest Ansermet, CH-1211 Geneva, Switzerland\label{Unige}
                \and Université Paris-Saclay, CNRS, Institut d’Astrophysique Spatiale, 91405, Orsay, France\label{IAS}
                \and University of Lyon, UCB Lyon 1, CNRS/IN2P3, IUF, IP2I Lyon, France\label{Lyon}
                \and Aix Marseille Univ, CNRS/IN2P3, CPPM, Marseille, France\label{CPPM}
                \and Department of Physics and Trottier Space Institute, McGill University, Montreal, QC, Canada H3A2T8\label{McGill}
                \and Institut de Recherche en Astrophysique et Planétologie (IRAP), Université de Toulouse, CNRS, UPS, CNES, 14 Av. Edouard Belin, 31400, Toulouse, France\label{IRAP}
                \\
                \email{fabien.lacasa@pd.infn.it}
        }
        
        \date{\today}
        
        \abstract
        { 
                Super-sample covariance (SSC) is an important effect for cosmological analyses that use the deep structure of the cosmic web; it may, however, be nontrivial to include it practically in a pipeline. We solve this difficulty by presenting a formula for the precision (inverse covariance) matrix and show applications to update likelihood or Fisher forecast pipelines. The formula has several advantages in terms of speed, reliability, stability, and ease of implementation. We present an analytical application to show the formal equivalence between three approaches to SSC: (i) at the usual covariance level, (ii) at the likelihood level, and (iii) with a quadratic estimator. We then present an application of this computationally efficient framework for studying the impact of inaccurate modelling of SSC responses for cosmological constraints from stage IV surveys. We find that a weak-lensing-only analysis is very sensitive to inaccurate modelling of the scale dependence of the response, which needs to be calibrated at the $\sim15\%$ level. The sensitivity to this scale dependence is less severe for the joint weak-lensing and galaxy clustering analysis (also known as 3x2pt). Nevertheless, we find that both the amplitude and scale-dependence of the responses have to be calibrated at better than 30\%.
        }
        
        \maketitle
        
        %%%%%%%%%%%%%%%%%%%%%%%%%%%%%%%%%%%%%%%%%%%%%%%%%%%%%%%%%%%%%%%%%%%
        
        \section{Introduction}\label{Sect:intro}
        
        Super-sample covariance (SSC) \citep{Hu2003,Takada2013}, also known as sample variance for cluster counts, is an important part of statistical errors or cosmic variance for the current survey of the large-scale structure (LSS) of the Universe \citep[e.g.][]{Hildebrandt2017}, and even more for near future cosmic surveys that aim to deepen our understanding of dark energy and gravity. Specifically, it is expected to be the dominant non-Gaussian contribution to the covariance of weak-lensing statistics \citep{Barreira2018b}, although other covariance terms are also important for galaxy clustering \citep{Lacasa2018b,Lacasa2020}. It has been shown that for a stage IV survey, the dark energy figure of merit can be degenerate by a factor of two \citep{Gouyou2022} when the SSC is accounted for in the joint analysis of weak-lensing (WL), photometric galaxy clustering (GC). This joint analysis is also called 3x2pt analysis.         
        However, it is difficult to estimate the SSC because it cannot be calibrated reliably with jackknife or bootstrap methods or simulations with sizes comparable to the survey size \citep{Lacasa2017}. 
        Instead, significant progress has recently been made in the analytical modelling of the SSC \citep[e.g.][]{Takada2014,Takahashi2014,Li2018,Chan2018,Lacasa2018,Wadekar2020} as well as in semi-analytical predictions of the SSC using measurements of the simulated reaction of observables to super-survey modes \citep[e.g.][]{Barreira2017a,Barreira2018}.
        An alternative approach that includes the effect of super-survey modes is to account for them with additional nuisance parameters that are marginalised over, that is, to account for them at the likelihood level and not in the covariance \citep{Li2014b,Lacasa2019}. In this article, we examine the two approaches to the modelling of the impact of super-survey modes (see Sect.~\ref{Sect:precision-matrix}). 
        
        We address four challenges in implementing the SSC in an analysis pipeline: change, speed, reliability, and stability. The first challenge, change, is the difficulty of changing an existing pipeline that may use assumptions that are broken by the SSC (e.g. that the covariance matrix is diagonal) and comparing new results with the previous pipeline version. The second challenge is that the covariance pipeline may be slowed down by the time necessary to predict the SSC and by the inversion of the total covariance matrix (if large), which is no longer diagonal. This challenge is not relevant when the covariance is fixed, but when it is varied with parameters and/or when the robustness of the results is tested with respect to changes in the covariance. The third challenge is about the stability of the results with respect to variations in the theoretical ingredients of the SSC prediction because the SSC literature is quite technical and evolves rapidly. The fourth challenge is a practical one found by the authors: the SSC gives an ill-conditioned covariance matrix so that the inversion becomes unstable when the other covariance term becomes negligible (e.g. when an analysis is pushed in the deep non-linear regime or in a combined probe case with large matrices).
        In the first section of this article, we propose a framework that solves these four problems. We then develop theoretical and numerical applications of the framework.
        
        In detail, the article is organised as follows. In Sect.~\ref{Sect:precision-matrix} we show that SSC can be computed as an update to the precision (i.e. inverse covariance) matrix, present examples for different probes, and show the application in likelihood or Fisher pipelines. In Sect.~\ref{Sect:equivalence} we use the previous formulae to show the equivalence of three approaches to SSC: at the covariance level, at the likelihood level, and with a quadratic estimator. In Sect.~\ref{Sect:application} we apply the framework to a mock stage IV photometric galaxy survey, for WL alone or in combination with GC. We conclude in Sect.~\ref{Sect:conclusion}.
        
        %%%%%%%%%%%%%%%%%%%%%%%%%%%%%%%%%%%%%%%%%%%%%%%%%%%%%%%%%%%%%%%%%%%
        
        \section{SSC as an update to the precision matrix}\label{Sect:precision-matrix}
        
        In this section, we show that the inverse covariance matrix can be computed as an update to the noSSC case in a numerically efficient manner. We then show applications to compute the likelihood or the Fisher matrix of an LSS observable.
        
        To do this, we use the Woodbury identity for matrix inversion \citep{Woodbury1950}. This identity states that if $A$ is an invertible $(n,n)$ matrix of known inverse that is updated with a matrix of the form $U C V,$ with $U, C,$ and $V$ the matrices of the respective shapes $(n,m)$, $(m,m),$ and $(m,n),$ and $C$ is invertible, then the inverse of the updated matrix is given by 
        \be \label{Eq:Woodbury-orig}
        \left(A + UCV \right)^{-1} = A^{-1} - A^{-1}U \left(C^{-1} + VA^{-1}U \right)^{-1} VA^{-1}.
        \ee
        More generally, even if $C$ is not invertible, we have
        \be \label{Eq:Woodbury-reg}
        \left(A + UCV \right)^{-1} = A^{-1} - A^{-1}U C\left(\mathds{1} + V A^{-1} U C \right)^{-1} VA^{-1}. 
        \ee
        We use the general form of the Woodbury identity given by Eq. \ref{Eq:Woodbury-reg} for the case that $A$ and $C$ are symmetric matrices and $V=U^T$. The original and updated matrices are therefore both symmetric.\\
        It is not immediately obvious that the right-hand side of Eq.~\ref{Eq:Woodbury-reg} is also symmetric, as it should be. However, this becomes clearer when $C$ is invertible, where we can use the equivalent form in Eq.~\ref{Eq:Woodbury-orig}, which shows that the right-hand side is symmetric. Then we can note that matrix symmetry is a continuous property, and invertible matrices are dense in matrix space. The property therefore holds by continuity even if $C$ is not invertible.
        
        %%%%%%%%%%%%%%%%%%%%
        \subsection{General formula}\label{Sect:formula}
        
        Let $\mathcal{O}(i,\alpha)$ be an abstract LSS probe, where $\alpha$ is the index (multi-index) for the redshift bin (redshift tuple), and $i$ is the index for all other quantities on which the observable depends (e.g. Fourier or angular wavevector, distance or angular separation, mass, size, or peak height bin). This observable is vectorised through an index $i_o$, with a mapping $i_o \rightarrow (i,\alpha)$ that is considered implicit hereafter.\\
        We call $n_z$ be the number of indices $\alpha$ and $n_o$ the number of indices $i_o$, that is, the size of the observable. For instance, if there is the same number $n_i$ of quantities for each redshift bin, then we have $n_o = n_i \times n_z$. This situation can, however, be broken by analysis choices employing a redshift-dependent range for the $i$ quantities, for example a cut $k_\mr{max}(z)$ or $\ell_\mr{max}(z)$.
        
        We use the Sij approximation to SSC (e.g. \citealt{Hu2003} for cluster counts, \citealt{Lacasa2019} for power spectra and other observables).
        This is a mean-field approximation for the redshift evolution, which is expected valid for observables with compact support in redshift and has been validated for use for cosmological analysis with clusters \citep{Hu2003,Aguena2018} or galaxies \citep{Lacasa2019}. In this approximation, the SSC of the probe is given by
        \be\label{Eq:SSC-base-eq}
        \Cov_\mr{SSC}\left[\mathcal{O}(i,\alpha) , \mathcal{O}(j,\beta) \right] = \partial_{\deltab} \mathcal{O}(i,\alpha) \ \partial_{\deltab} \mathcal{O}(j,\beta) \ S_{\alpha,\beta},
        \ee
        where $\deltab = (\rho_\mr{survey}-\bar{\rho})/\bar{\rho}$ is the change in matter density in the survey with respect to the global mean. This is also known as the super-survey mode. The $S$ matrix is the covariance matrix of these $\delta_b$ in each redshift bin, due to the clustering of matter. $\partial_{\deltab} \mathcal{O}$ is the reaction of the probe $\delta_b$, also known as the response. In practice, this response is computed either from a nonlinear theory \citep[e.g.][]{Takada2013,Lacasa2016}, from an ansatz \citep{Lacasa2019}, or from simulations \citep{Barreira2018}.\\
        In the following we often shorten $\mathcal{C}_\mr{SSC} \equiv \Cov_\mr{SSC}$, that is,
        \ba
        \nonumber\left(\mathcal{C}_\mr{SSC}\right)_{i_o,j_o} &= \Cov_\mr{SSC}\left[\mathcal{O}(i_o) , \mathcal{O}(j_o) \right] \\
        &= \Cov_\mr{SSC}\left[\mathcal{O}(i,\alpha) , \mathcal{O}(j,\beta) \right],
        \ea
        where $i_o \rightarrow (i,\alpha)$ and $j_o \rightarrow (j,\beta)$ through the implicit mapping mentioned previously.
        
        We now remark that this SSC matrix can be cast in the following form:
        \be\label{Eq:SSC-USU_form}
        \mathcal{C}_\mr{SSC} = U \, S \, U^T,
        \ee
        where $S$ is a square matrix of shape $(n_z,n_z)$ (making it evident that $\mathcal{C}_\mr{SSC}$ has rank $\leq n_z$), and $U$ is a rectangular matrix of shape $(n_o,n_z)$ given by
        \be\label{Eq:def-U}
        U_{i_o,\beta} = \partial_{\deltab} \mathcal{O}(i,\alpha) \ \delta_{\alpha,\beta},
        \ee
        where the final $\delta$ on the right-hand side is a Kronecker symbol. 
        
        Therefore, using the Woodbury identity in Eq.~\ref{Eq:Woodbury-reg}, the inverse of the total covariance matrix $\mathcal{C}_\mr{tot} = \mathcal{C}_\mr{noSSC} + \mathcal{C}_\mr{SSC}$ is given by
        \ba\label{Eq:precision-tot}
        \nonumber \mathcal{C}_\mr{tot}^{-1} &= \mathcal{C}_\mr{noSSC}^{-1} \\
        & - \mathcal{C}_\mr{noSSC}^{-1} \, U \, S \, \left(\mathds{1}+I(\deltab) \, S \right)^{-1}  U^T \, \mathcal{C}_\mr{noSSC}^{-1},
        \ea
        where we introduced the notation (that we justify below)
        \ba\label{Eq:def-Idb}
        I(\deltab) \equiv U^T \, \mathcal{C}_\mr{noSSC}^{-1} \, U.
        \ea
        Formula Eq.~\ref{Eq:precision-tot}  thus gives the total precision matrix as an update to the noSSC case. It is particularly useful numerically when the SSC dominates the covariance matrix elements: the SSC matrix indeed has a low rank and will thus destabilise the covariance inversion. This was, for instance, noted previously by one of the authors in numerical investigations pushing angular power spectra analyses at high multipoles in the SSC-dominated regime \citep{Lacasa2020}.
        
        %%%%%%%%%%%%%%%%%%%%
        \subsection{Examples}
        
        We now apply the general formula described in the previous subsection to specific class of LSS observables.
        The case of one-point functions is described in Sect.~\ref{Sect:ex-1pt}, the case of two-point functions is set out in Sect.~\ref{Sect:ex-2pt}, and the case of other statistics is presented in Sect.~\ref{Sect:ex-otherstats}.
        
        %%%%%%%%%%
        \subsubsection{Counts or one-point functions}\label{Sect:ex-1pt}
        The first class of observables includes counts of clusters as a function of (bins of) mass proxy (and redshift), of voids as a function of size, or of WL peaks as a function of peak height. In all cases, the observable can be noted $N_\mr{obj}(i,\alpha)$: the number of objects as a function of the index $i$ of quantity (mass, size, and height) and redshift index $\alpha$. For clusters and voids, and except for factors of confusion in bin assignment (e.g. mass errors or photo-z errors), the noSSC (=standard) covariance is Poissonian,
        \ba
        \Cov_\mr{noSSC}\left[N_\mr{obj}(i,\alpha) , N_\mr{obj}(j,\beta) \right] = N_\mr{obj}(i,\alpha) \, \delta_{i,j} \, \delta_{\alpha,\beta},
        \ea
        where the final $\delta$ on the right-hand side are Kronecker symbols, that is, the noSSC matrix is diagonal. This diagonality somewhat simplifies Eq.~\ref{Eq:precision-tot},
        \ba
        \nonumber \mathcal{C}_\mr{tot}^{-1} &\left[ N_\mr{obj}(i,\alpha) , N_\mr{obj}(j,\beta) \right] = N_\mr{obj}^{-1}(i,\alpha) \, \delta_{i,j} \, \delta_{\alpha,\beta} \\
        \nonumber - & \sum_\gamma \left[\partial_{\deltab} N_\mr{obj}(i,\alpha)\right] N_\mr{obj}^{-1}(i,\alpha) \times S_{\alpha,\gamma} \times \left(\mathds{1}+I(\deltab) \, S\right)^{-1}_{\gamma,\beta} \\
        & \times \left[\partial_{\deltab} N_\mr{obj}(j,\beta)\right] N_\mr{obj}^{-1}(j,\beta),
        \label{Eq:precision-counts-expanded}
        \ea
        which can be written in a more compact form as
        \ba
        \nonumber \mathcal{C}_\mr{tot}^{-1} &\left[(i,\alpha) , (j,\beta) \right] =  \frac{\delta_{i,j} \, \delta_{\alpha,\beta}}{N_\mr{obj}(i,\alpha)} \\
        - & \ \partial_{\deltab}\! \ln N_\mr{obj}(i,\alpha) \left(S \, (\mathds{1}+I(\deltab) \, S)^{-1}\right)_{\alpha,\beta} \;\! \partial_{\deltab}\! \ln N_\mr{obj}(j,\beta).
        \label{Eq:precision-counts-compact}
        \ea
        Here, the matrix $I(\deltab)$ has the shape $(n_z,n_z)$, and is given by
        \ba
        I(\deltab)_{\alpha,\beta} = \left(\sum_i \left[\partial_{\deltab} N_\mr{obj}(i,\alpha)\right]^2 \ N_\mr{obj}^{-1}(i,\alpha)\right) \delta_{\alpha,\beta}.
        \ea
        We note that $\mathds{1}+I(\deltab) \, S$ is not symmetric (but $S \cdot (\mathds{1}+I(\deltab) \, S)^{-1}$ is) and must be inverted numerically. This is not a problem because (i) it is regular (even if $S$ is not invertible, the total is regularised by the presence of the identity matrix), and (ii) it has a small size: the number of redshift bins. 
        
        %%%%%%%%%%
        \subsubsection{Power spectra or two-point functions}\label{Sect:ex-2pt}
        The second class of observable includes the (angular) power spectra or the (angular) correlation function of galaxies, weak-lensing, clusters, cosmic microwave background lensing, HI intensity mapping, and so on as a function of Fourier (harmonic) wave number or (angular) separation and of redshift bin. If we only consider auto-spectra (i.e. correlate only in the same redshift bin), then $\alpha$ is the index of the redshift bin. If we also consider cross-spectra, then $\alpha$ is the index of the redshift pair. Even in the noSSC auto-spectrum case, there is often cross-talk between redshift bins, for example because of photo-z errors, so that the covariance is not diagonal in this direction.
        However, the noSSC covariance is diagonal in terms of Fourier or harmonic wave number under two conditions: (i) it may be argued that the Gaussian covariance is the dominant term beyond SSC (\cite{Barreira2018b} showed this is the case for WL, although \cite{Lacasa2020} reported that other non-Gaussian terms are important for constraints on $n_S$ with GC.), and (ii) the wave numbers are binned sufficiently to erase the mode-coupling due to the survey mask \citep[see e.g.][]{Hivon2002}.
        As for counts, this slightly simplifies Eq.~\ref{Eq:precision-tot}, although less so because we still have redshift correlations. For example, for an angular power spectrum $C_\ell(\alpha)$, the Gaussian covariance is a bloc diagonal multipole by multipole. We call $\Sigma^2_\ell$ the multipole bloc with the shape $(n_z,n_z)$, 
        \be
        \Sigma^2_\ell(\alpha,\beta) = \Cov\left(C_\ell(\alpha),C_\ell(\beta)\right),
        \ee
        and denote its inverse with $\Sigma^{-2}_\ell$. We thus have
        \ba
        \nonumber \mathcal{C}_\mr{tot}^{-1} &\left[C_\ell(\alpha) , C_{\ell'}(\beta) \right] = \Sigma^{-2}_\ell(\alpha,\beta) \ \delta_{\ell,\ell'} \\
        \nonumber - & \sum_{\gamma,\epsilon,\zeta} \partial_{\deltab} C_\ell(\gamma) \  \Sigma^{-2}_\ell(\gamma,\alpha) \ S_{\gamma,\epsilon} \ (\mathds{1}+I(\deltab) \, S)^{-1}_{\epsilon,\zeta} \\
        & \times \partial_{\deltab} C_{\ell'}(\zeta) \ \Sigma^{-2}_\ell(\zeta,\beta),
        \label{Eq:precision-cl-expanded}
        \ea
        which can be written in a more compact form as
        \ba
        \mathcal{C}_\mr{tot}^{-1} =  \mathcal{C}_\mr{Gauss}^{-1} - N_\ell^T \, S \, (\mathds{1}+I(\deltab) \, S)^{-1} N_{\ell'}, \label{Eq:precision-cl-compact}
        \ea
        where the matrix $N_\ell$ has the shape $(n_z,n_z)$ and we have
        \ba
        N_\ell(\alpha,\beta) = \partial_{\deltab} C_\ell(\alpha) \ \Sigma^{-2}_\ell(\alpha,\beta) .
        \ea
        Here, the matrix $I(\deltab)$ has the shape $(n_z,n_z)$ and is given by
        \ba
        I(\deltab)_{\alpha,\beta} = \sum_{\ell} \partial_{\deltab} C_\ell(\alpha) \, \Sigma^{-2}_\ell(\alpha,\beta) \, \partial_{\deltab} C_\ell(\beta) .
        \ea
        
         Equation~\ref{Eq:precision-cl-compact} greatly decreases the numerical cost of inverting $\mathcal{C}_\mr{tot}$. For the covariance matrix of shape $(n_\ell \times n_z,n_\ell \times n_z)$, the cost of brute force computation is $\mathcal{O}((n_\ell \times n_z)^3)$, while through Eq.~\ref{Eq:precision-cl-compact}, we invert only (many) $(n_z,n_z)$ matrices, reducing the cost to $\mathcal{O}(n_\ell \times n_z^3)$. Therefore, we achieve a speedup factor of $n_\ell^2$. For a typical analysis with tens of multipole bins (e.g. we use 40 bins in Sect.~\ref{Sect:application}), this results in an appreciable speedup factor in the hundreds. This speedup means that a larger number of multipole bins can be used instead of  binning aggressively to extract as much information as possible from the measurement, for instance, to hunt for localised features.
        
        %%%%%%%%%%
        \subsubsection{Other statistics}\label{Sect:ex-otherstats}
        The formulae can easily be generalised to other statistics. For instance, a combination of angular power spectrum measurements of different probes amounts to generalising the $\alpha$ multi-index to index both probes and redshift tuples. We use this in Sect.~\ref{Sect:application} for the so-called 3x2pt analysis, which combines WL, GC, and their cross correlation (XC). Higher-order statistics are also possible. For instance, for a bispectrum analysis, $i$ would become a multi-index for triplets of wavenumbers, and $\alpha$ becomes a multi-index for triplets of redshift bins. 
        
        In all cases, the formula becomes particularly useful when the noSSC covariance matrix is diagonal in some direction, as it can leverage this to lower the numerical cost of the inversion. For instance, for a combination of power spectra, the Gaussian covariance only correlates with identical wavenumbers, and for bispectra, it only correlates with identical wavenumber triplets.
        
        %%%%%%%%%%%%%%%%%%%%
        \subsection{Applications}\label{Sect:update-applications}
        
        %%%%%%%%%%
        \subsubsection{Log likelihood}\label{Sect:loglike}
        For a standard Gaussian likelihood with fixed covariance, we have
        \ba
        -2 \ln\mathcal{L} = (d-m)^T \ \mathcal{C}_\mr{tot}^{-1} (d-m),
        \ea
        where $d$ is the data vector and $m$ is the corresponding model.\\
        Adding SSC can thus be considered as an update to the log-likelihood: $\ln\mathcal{L}=\ln\mathcal{L}_\mr{noSSC}+\delta\ln\mathcal{L}$ with
        \ba\label{Eq:update-loglike}
        -2 \ \delta\ln\mathcal{L} = X^T \, S \, (\mathds{1}+I(\deltab) \, S)^{-1} \, X,
        \ea
        where $X = U^T \, \mathcal{C}_\mr{noSSC}^{-1} \, (d-m)$. This makes it easy and numerically efficient to update a noSSC pipeline to account for SSC. In the noSSC pipeline, the covariance matrix $\mathcal{C}_\mr{noSSC}$ should already have been inverted, therefore no numerical cost is added there. Then we have a matrix operation to downsize the problem into multiplication and inversion of small $(n_z,n_z)$ matrices, which is cheap with respect to all the other operations. Hence the computation of $\delta\ln\mathcal{L}$ takes a time comparable to that of $\ln\mathcal{L}_\mr{noSSC}$ and is not a bottleneck of the pipeline.
        
        If the covariance is allowed to vary with model parameters, there is an additional determinant term in the likelihood. The impact of SSC can be computed through the determinant lemma,
        \ba
        \det\left(A + U W V^T\right) = \det\left(\mathds{1} + V^T A^{-1} U W\right) \times \det\left(A\right),
        \ea
        which gives
        \ba
        \ln\det\mathcal{C}_\mr{tot} = \ln\det\mathcal{C}_\mr{noSSC} + \ln\det\left(\mathds{1} + U^T \mathcal{C}_\mr{noSSC}^{-1} U S\right).
        \ea
        We thus keep the property $\ln\mathcal{L}=\ln\mathcal{L}_\mr{noSSC}+\delta\ln\mathcal{L}$, and the additive determinant term is numerically cheap since it is the determinant of a $(n_z,n_z)$ matrix.
        
        %%%%%%%%%%
        \subsubsection{Fisher matrix}\label{Sect:Fisher}
        We consider the observable $\mathcal{O}$ as a vector. The vector of the model parameters is $\theta$ with a length $n_\theta$, and the parameters are indexed with Greek indices $\mu,\nu,\ldots$. With these notations, the Fisher matrix elements are given by
        \ba
        F_{\mu,\nu} = \partial_\mu \mathcal{O}^T \, \mathcal{C}_{tot}^{-1} \, \partial_\nu \mathcal{O},
        \ea
        where $\partial_\mu \mathcal{O}$ is the partial derivative of the  observable with respect to the parameter $\mu$, considered as a vector. If we plug in Eq.~\ref{Eq:precision-tot}, this yields
        \ba\label{Eq:Fisher-multibins-munu}
        F_{\mu,\nu} = F_{\mu,\nu}^\mr{noSSC} - Y_\mu^T \ S \ (\mathds{1}+I(\deltab) \, S)^{-1} \ Y_\nu,
        \ea
        where $Y_\mu = U^T \, \mathcal{C}_\mr{noSSC}^{-1} \, \partial_\mu \mathcal{O}$ is a vector with the size $n_z$. We can thus consider $Y$ as a $(n_z,n_\theta)$ matrix, which allows rewriting Eq.~\ref{Eq:Fisher-multibins-munu} in matrix form,
        \ba\label{Eq:Fisher-multibins}
        F = F_\mr{noSSC} - Y^T \ S \ \left(\mathds{1}+I(\deltab) \ S\right)^{-1} \ Y.
        \ea
        In particular, for a single redshift bin ($n_z=1$), we obtain
        \ba\label{Eq:Fisher-singlebin}
        F = F_\mr{noSSC} - \frac{Y^T \ Y \ S}{1+ I(\deltab) S},
        \ea
        where $Y$ is a covector and $I(\deltab) = \partial_{\deltab} \mathcal{O}^T \, \mathcal{C}_\mr{noSSC}^{-1} \, \partial_{\deltab} \mathcal{O}$ is a scalar that can be interpreted as the Fisher information on \deltab as an additional parameter. This interpretation will become clearer in Sect.~\ref{Sect:equivalence}. We note that Eq.~\ref{Eq:Fisher-singlebin} recovers a result from \cite{Lacasa2019}, specifically, their Eq.~15 in matrix form, with the mapping $Y_\mu \leftrightarrow f_\mu^\mr{SSC}$, $I(\deltab) \leftrightarrow V^T \, \mathcal{C}_\mr{noSSC}^{-1} \, V$, and $S \leftrightarrow S_{i,i}$. We note, however, that the latter result is limited to a single redshift bin (which is admittedly of little practical interest), while the general result in Eq.~\ref{Eq:Fisher-multibins-munu} applies to any observable(s) in any number of bins. 
        
        In the multi-bin case, $I(\deltab)$ is a $(n_z,n_z)$ matrix that represents the Fisher information on a vector of additional parameters $\deltab(\alpha)$, indeed we have
        \ba
        I(\delta_b)_{\alpha,\beta} = \sum_{i,j} \partial_{\deltab} \mathcal{O}(i,\alpha) \ \mathcal{C}_\mr{noSSC}^{-1}(i,\alpha;j,\beta) \ \partial_{\deltab} \mathcal{O}(j,\beta).
        \ea
        
        %%%%%%%%%%
        \subsubsection{Inverse Fisher matrix}\label{Sect:iFisher}
        The inverse of the Fisher matrix is required to forecast marginalised errors on the parameters. Fortunately, the form of Eq.~\ref{Eq:Fisher-multibins} means that the Woodbury inversion formula can be used again. After some algebra, this yields
        \ba
        F^{-1} &= F^{-1}_\mr{noSSC} + \delta F^{-1}, \\
        \delta F^{-1} &= F_\mr{noSSC}^{-1} \, Y^T \, S \, (\mathds{1}+(I(\delta_b) - \delta I) \, S)^{-1} Y \, F_\mr{noSSC}^{-1},
        \ea
        with
        \ba\label{Eq:def-dI}
        \delta I = Y \, F_\mr{noSSC}^{-1} \, Y^T.
        \ea
        This can be rewritten in the form that is used throughout this article:
        \ba\label{Eq:iFisher-multibins}
        F^{-1} &= F^{-1}_\mr{noSSC} + Z^T \ \Sigma^2(\deltab) \ Z,
        \ea
        with
        \ba
        Z &\equiv Y \, F_\mr{noSSC}^{-1} \label{Eq:def-Z}, \\
        \Sigma^2(\deltab) &\equiv S \, (\mathds{1}+(I(\delta_b) - \delta I) \, S)^{-1},
        \label{Eq:def-sigma2deltab}
        \ea
        whose physical interpretations will become clearer later.
        
        Eq.~\ref{Eq:iFisher-multibins} allows forecasting the impact of SSC on the parameter constraints as an update to the noSSC case. This has advantages in terms of speed for two reasons: (i) The noSSC pipeline does not need to be rewritten nor rerun, its results can be reused, and (ii) if an assumption of diagonality of the noSSC covariance is used, it can also be used for the SSC Fisher elements: it will speed up the computation of $Y$, and from then on, only cheap matrix multiplications are necessary (matrix sizes: $n_z$ or $n_\theta$).
        
        %%%%%%%%%%%%%%%%%%%%%%%%%%%%%%%%%%%%%%%%%%%%%%%%%%%%%%%%%%%%%%%%%%%
        
        \section{Equivalence of SSC approaches}\label{Sect:equivalence}
        
        In this section, we show that three approaches to SSC are equivalent under the conditions detailed in the respective subsections. The first is the classical approach used in the previous section, where SSC is included as an additional covariance term. In the second approach, SSC is instead accounted for by promoting \deltab to be a nuisance parameter (single bin case; \cite{Li2014b}) or vector of parameters (one per bin, multi-bin case; \cite{Lacasa2019}) that is marginalised over with a prior $\mathcal{N}(0,S)$. The third approach is new. It estimates \deltab with a quadratic estimator (QE) to undo the SSC effect.
        
        %%%%%%%%%%%%%%%%%%%%
        \subsection{Equivalence between the covariance and likelihood approaches in the single bin case}\label{Sect:equiv-cov-like}
        
        In the case of a single redshift bin, the inverse Fisher matrix in the covariance approach is a special case of Eq.~\ref{Eq:iFisher-multibins},        \ba\label{Eq:iFisher-singlebin}
        F^{-1} = F^{-1}_\mr{noSSC} + \frac{Z^T \, Z \ S}{1+(I(\deltab)-\delta I)S}. 
        \ea
        
        In the likelihood approach, the covariance of the data is the noSSC covariance, and the parameter vector has been extended to include \deltab: $\theta' = (\theta,\deltab)$. This gives a Fisher matrix,
        \ba\label{Eq:Fisherlike-freedb}
        F_{\mu,\nu}(\mr{free \ \deltab}) = \partial_\mu \mathcal{O}^T \cdot \mathcal{C}_{\mr{noSSC}}^{-1} \cdot \partial_\nu. \mathcal{O},
        \ea
        where $\mu$ and $\nu$ are the indices of the model parameters, running from 1 to $n_\theta$.\\
        On top of this, a prior $\mathcal{N}(0,S)$ is imposed on $\deltab$. Because the Fisher matrix is the inverse covariance of the parameters and because inverse covariances of independent Gaussians add up, imposing this prior is equivalent to adding a Fisher matrix, 
        \ba
        F_{\mu,\nu}(\mr{prior}) = \frac{1}{S} \ \delta_{\deltab,\mu} \ \delta_{\deltab,\nu},
        \ea
        where the $\delta$ on the right-hand side are Kronecker symbols.
        
        When we introduce the notations of Sect.~\ref{Sect:Fisher}, the total Fisher matrix $F'=F(\mr{free \ \deltab})+F(\mr{prior})$ then takes the form
        \ba\label{Eq:Fisher-tot-likelihoodapproach}
        F' = \begin{pmatrix} F^\mr{noSSC} & Y^T \\ Y & I(\deltab)\end{pmatrix} + \begin{pmatrix} 0 & 0 \\ 0 & \frac{1}{S} \end{pmatrix}.
        \ea
        Then some algebra that is detailed in Appendix \ref{App:iFisher-lik} for transparency shows that the inverse Fisher matrix can be written as
        \ba \label{Eq:F'm1-finalform}
        F'^{-1} &= \begin{pmatrix} F^{-1}_{\mr{noSSC}} & 0 \\ 0 & 0\end{pmatrix} + \frac{S}{1 + (I-\delta I) S} \begin{pmatrix} Z^T Z & Z^T \\ Z & 1\end{pmatrix}.
        \ea
        In particular, for the lower right element, we have
        \ba \label{Eq:F'm1-deltab}
        F'^{-1}_{\deltab,\deltab} = \frac{S}{1 + (I-\delta I) S}.
        \ea
        The upper left block for the parameters $\theta$ reads
        \ba\label{Eq:F'm1-theta}
        F'^{-1} = F^{-1}_{\mr{noSSC}} + \frac{S}{1 + I S - \delta I \, S} Z^T Z.
        \ea
        It is identical to Eq.~\ref{Eq:iFisher-singlebin}, which shows that the covariance and likelihood approaches give the same constraints on the model parameters, that is, they are equivalent.
        
        It seems natural that the equivalence should extend to the case of several bins, although we did not attempt this calculation.
        
        %%%%%%%%%%%%%%%%%%%%
        \subsection{Equivalence between the QE and covariance approaches for the power spectrum on the same multipole range}
        
        In this section, we denote estimators with an upper hat. For example, $\hat{C}_\ell$ is the angular power spectrum estimator. The expectation value is noted with an upper bar, for instance $\overline{C}_\ell$, and can be predicted with a cosmological code as a function of $\theta$ the vector of the model parameters. We assume the theoretical predictions to be unbiased.\\
        A QE can be built to estimate the large-scale modes of the density field using their coupling to better measure small-scale modes (see e.g. \cite{Li2020,Darwish2020}). The natural idea is to use such an estimator to measure \deltab , which is the zero mode of the survey, and then combine this with the traditional power spectrum measurement. However, for this zero mode, the QE collapses to a linear combination of power spectrum measurements in full sky. For instance, in terms of angular observables and using notations from the QE literature, Appendix \ref{App:QE} shows that in the full sky, the \deltab estimator is
        \be \label{Eq:def-QE}
        \hat{\deltab} = \frac{1}{N} \sum_{\ell} \ g_{\ell,\ell}(0) \, (\hat{C}_\ell-\overline{C}_\ell(\theta)),
        \ee
        with 
        \be \label{Eq:norm-QE}
        N = \sum_{\ell,\ell'} \partial_{\deltab} C_\ell \ \mathcal{C}^{-1}_{\mr{noSSC};\ell,\ell'} \ \partial_{\deltab} C_{\ell'},
        \ee
        and 
        \be
        g_{\ell,\ell}(0) = \sum_{\ell'} \mathcal{C}^{-1}_{\mr{noSSC};\ell,\ell'} \ \partial_{\deltab} C_{\ell'}.
        \ee
        This is consistent with comparable equations from \cite{Li2014b}, who worked on 3D observables and found an estimator written in terms of $P(k)$ measurements.
        
        If the QE and the power spectrum are estimated on the same field, and if the multipole range used in the QE is included in the multipole range in which the power spectrum is measured, then the QE measurement is entirely determined by the power spectrum measurement. The joint probability is
 accordingly        \ba
        P(\hat{C},\hat{\deltab}) = P(\hat{C}) \times \delta_D \left(\hat{\deltab} - \frac{1}{N} \sum_{\ell} \ g_{\ell,\ell}(0) \, (\hat{C}_\ell-\overline{C}_\ell)\right),
        \ea
        where $\overline{C}_\ell$ is fixed at its value in the fiducial cosmology needed to build the QE. As in Sect. \ref{Sect:equiv-cov-like}, we note $\theta'=(\theta,\deltab)$ the vector of parameters. The likelihood then reduces to 
        \ba
        \nonumber -\ln \mathcal{L}(\hat{C},\hat{\deltab}|\theta') &= -\ln \mathcal{L}(\hat{C}|\theta') \\
        & - \ln \delta_D\left(\hat{\deltab}  - \frac{1}{N} \sum_{\ell} \ g_{\ell,\ell}(0) \, (\hat{C}_\ell-\overline{C}_\ell)\right),
        \ea
        where the first term is the usual likelihood from the power spectrum alone. In the second term, $\delta_D$ is a Dirac distribution. This second term has no dependence on the model parameters (e.g. for the weights $g_{\ell,\ell}(0)$, all cosmological parameters are fixed at the fiducial value used to build the estimator). The Fisher matrix is given in terms of the likelihood by 
        \ba
        F_{\alpha,\beta} = \lbra -\frac{\partial^2 \ln \mathcal{L}}{\partial \theta'_{\alpha} \, \partial \theta'_{\beta}} \middle| \theta' \rbra,
        \ea
        where $\lbra \middle| \theta'\rbra$ denotes the conditional expectation given the values of $\theta'$.\\
        If we call $F''$ the Fisher matrix of this QE + power spectrum approach, we obtain
        \ba
        \nonumber F''_{\alpha,\beta} &= F'_{\alpha,\beta} + \lbra -\frac{\partial^2 \ln \delta_D\left(\hat{\deltab} - \cdots\right)}{\partial \theta'_{\alpha} \, \partial \theta'_{\beta}} \middle| \theta'\rbra, \\
        &= F'_{\alpha,\beta},
        \ea
        where the second term vanished because the derivatives with respect to model parameters vanish. Thus we have the same Fisher constraints as the likelihood of the power spectrum, as well as the traditional covariance approach for constraints on $\theta$.
        
        %%%%%%%%%%%%%%%%%%%%
        \subsection{Physical interpretation and conclusion}
        
        In this section, we use the two previous sections to give a physical interpretation to some of the quantities introduced in Sects.~\ref{Sect:Fisher} and \ref{Sect:iFisher} : $I(\deltab)$, $\Sigma^2(\deltab)$, $\delta I,$ and $Z$.
        
        First, Eq.~\ref{Eq:Fisher-tot-likelihoodapproach} shows that $I(\deltab)$ is the direct Fisher information on $\deltab$. Furthermore, from the QE normalisation in Eq. \ref{Eq:norm-QE}, we can recognise that $N = I(\deltab)$. Computing the variance of the QE gives
        \be
        <\hat{\deltab}^2\!\!> = \frac{1}{N} = \frac{1}{I(\deltab)}.
        \ee
        Hence $I(\deltab)$ is the inverse variance of the maximum likelihood estimator for $\deltab$, which consolidates its interpretation as the raw information on \deltab when all other model parameters are fixed.
        
        Second, $\Sigma^2(\deltab)$ is defined in Eq.~\ref{Eq:def-sigma2deltab}, and Eq.~\ref{Eq:F'm1-deltab} (in the single bin case) shows that it is the marginalised Fisher constraint on $\deltab$. In the single bin case, it can be cast in the form
        \be
        \Sigma^2(\deltab) = \frac{1}{I(\deltab) - \delta I + \frac{1}{S}},
        \ee
        which makes it clear that there is a difference with the unmarginalised QE error due to (i) the presence of a prior $\frac{1}{S}$ , which adds information, (ii) the negative term $- \delta I$ , which removes information. In the limit of no prior, we have $S\rightarrow\infty$ and $\Sigma^2(\deltab) \rightarrow \frac{1}{I(\deltab) - \delta I}$. We therefore conclude that $\delta I$ quantifies the loss of information on \deltab due to the degeneracy with the other model parameters in $\theta$.
        
        Third, the off-diagonal matrix elements of Eq.~\ref{Eq:F'm1-finalform} show that the covector $Z$ quantifies the degeneracy between \deltab and each model parameter in $\theta$. Furthermore, when we write Eq.~\ref{Eq:F'm1-theta} on the diagonal, we have
        \be
        F^{-1}_{\mu,\mu} = F^{-1}_{\mr{noSSC},\mu,\mu} + \Sigma^2(\deltab) \left(Z_\mu\right)^2,
        \ee
        which shows that $Z_\mu$ is the (square root of the) increase in variance on parameter $\mu$ as a function of the constraint on $\deltab$. To reflect this, we could have noted this quantity $\sqrt{\frac{\partial \sigma^2(\mu)}{\partial \sigma^2(\deltab)}}$ or perhaps $\delta \sigma_\mr{rms}^\mr{SSC}(\mu)$ , but this would have made the notations of all previous equations in Sect.~\ref{Sect:equivalence} much heavier and harder to read. We therefore chose to keep a simple $Z$.
        
        Finally, as a summary of this section, we have shown that accounting for SSC at the likelihood level is formally equivalent, in terms of parameter constraints, to accounting for it traditionally at the covariance level. Combining a QE and a power spectrum does not bring better parameter constraints than the power spectrum alone as long as both estimators act upon the same field and the same multipole range.\\
        The  $\hat{\deltab}$ estimator might be used mainly (i) when some observational effects can be dealt with more efficiently (e.g. survey mask or systematics) in the QE than in the power spectrum analysis, (ii) if it allows adding information from another field (e.g. estimate \deltab with galaxies to improve a WL analysis), from another statistic (e.g. a linear, cubic, or log-normal estimator that uses information from the one-point or three-point function or from the log field), or from another survey \citep[see e.g.][]{Digman2019} when a full combined probe analysis would be too costly.
        
        %%%%%%%%%%%%%%%%%%%%%%%%%%%%%%%%%%%%%%%%%%%%%%%%%%%%%%%%%%%%%%%%%%%
        
        \section{Application to a stage IV weak-lensing and galaxy clustering analysis}\label{Sect:application}
        
        In this section, we apply the previous results to the case of a future stage IV photometric galaxy survey that constrains cosmology through WL alone and in combination with GC and XC. We describe the experimental and modelling setup in Sect.~\ref{Sect:setup}. In Sect.~\ref{Sect:info-deltab} we examine how well the survey can constrain the super-sample modes. In Sect.~\ref{Sect:asymto-errors} we ask whether any theoretical prior on $\deltab$ can be lifted. Finally, in Sect.~\ref{Sect:requirements} we study the precision to which the SSC responses should be modelled in order to pass certain precision requirements on the reported cosmological errors.
        
        %%%%%%%%%%%%%%%%%%%%
        \subsection{Setup}\label{Sect:setup}
        
        For the cosmological modelling, we worked within the $w_0 w_a$ CDM model, and in the Fisher forecast, we varied each of its parameters: $\Omega_m$, $\Omega_b$, $w_0$, $w_a$, $h$, $n_S,$ and $\sigma_8$. In practice, we mostly focused on the parameters that are most affected by SSC: $w_0$, $w_a$, $\sigma_8,$ and $\Omega_m$. For the prediction of all cosmological quantities, we used the software \texttt{cosmosis} by \cite{Zuntz2015}, and for the prediction of the matter power spectrum $P(k)$ on non-linear scales, we used the prescription by \cite{Takahashi2012} and \cite{Bird2012}. 
        
        In terms of experimental specifications, we followed all the specifications of \cite{Euclid-IST} for a 15'000 deg$^2$ photometric galaxy survey to analyse WL, GC and XC.
        In slightly more detail, the galaxy redshift distribution is
        \be
        n(z) \propto \left(\frac{z}{z_0}\right)^2 \, \exp\left[-\left(\frac{z}{z_0}\right)^{3/2}\right]\,,
        \ee
        with $z_0=0.9/\sqrt{2}$ and a total density of 30 galaxies per square ${\rm arcmin}$. Then we defined ten tomographic redshift bins such that they are equi-populated, yielding the boundaries $z_i=(0.0010,0.42,0.56,0.68,0.79,0.90,1.02,1.15,1.32,1.58,2.50)$. The redshifts were then perturbed by photometric errors with the main dispersion and a population of outliers, following the specific details given in \cite{Euclid-IST}. These galaxies are both the source and lens sample, that is, their shape was used for the WL measurement and their position for the GC measurement.
        
        In terms of probe modelling, we also followed \cite{Euclid-IST}. The GC was modelled with a scale-independent bias with a fiducial value $b(z)=\sqrt{1+z}$. For cosmological constraints, the galaxy bias value in each tomographic bin is a nuisance parameter that needs to be marginalised over. The WL is modelled classically as a tracer of matter, its power is hampered on small scales by the shape noise with standard deviation $\sigma_{\epsilon}=0.3$. Another issue for WL is the contribution from the intrinsic alignments (IA) of galaxies. To model this effect, we followed \cite{Euclid-IST}: the IA power spectrum and cross-spectrum with matter are given by $P_\mr{II} = [-A(z)]^2 P_\mr{mm}$ and $P_\mr{mI} = -A(z) P_\mr{mm}$, where
        \be
        A(z) = \mathcal{A}_\mr{IA} \frac{ \mathcal{C}_\mr{IA} \, \Omega_m \, \mathcal{F}_{\rm IA}(z)}{D(z)},
        \ee
        with $\mathcal{C}_\mr{IA}=0.0134$ being a normalisation constant, $D(z)$ the growth factor, and $\mathcal{A}_\mr{IA}$ a nuisance parameter that controls the amplitude of the IA contribution. The redshift dependence is given by
        \be
        \mathcal{F}_\mr{IA}(z) =(1+z)^{\eta_\mr{IA}} \left[\frac{\braket{L}(z)}{L_*(z)}\right]^{\beta_\mr{IA}},
        \ee
        where $\braket{L}(z)/L_*(z)$ is the ratio of the mean source luminosity and the characteristic scale of the luminosity function \citep{Hirata2007,Bridle2007}. Additionally, we have two nuisance parameters, $\eta_\mr{IA}$ and $\beta_\mr{IA}$, making a total of three nuisance parameters for intrinsic alignments, with fiducial values \citep{Euclid-IST} $(\mathcal{A}_\mr{IA},\eta_\mr{IA},\beta_\mr{IA}) = (1.72, -0.41, 2.17)$.
        
        For the statistical observable, we used the angular power spectrum for all probes: the WL auto-spectrum, the GC auto-spectrum, and the XC cross-spectrum. For simplicity, we adopted a common range of multipoles from $\ell_\mr{min}=10$ to $\ell_\mr{max}=3000$ for all probes. The covariance of these power spectra is described by a classical Gaussian term (including shape noise for WL and shot-noise for GC), and the super-sample covariance term is accounted for through the developments of Sect.~\ref{Sect:precision-matrix}. Specifically for the SSC, we computed the $S$ matrices using \href{https://pyssc.readthedocs.io}{\texttt{PySSC}} \citep{Lacasa2019}, and for the responses, we followed \cite{Gouyou2022} and took $\partial_{\deltab} C_\ell = R \times C_\ell$ with a probe-independent dimensionless response $R=4$. Following \cite{Barreira2018b}, other non-Gaussian covariance terms have a negligible impact on cosmological constraints from a WL survey, hence we neglected these terms here.
        
        %%%%%%%%%%%%%%%%%%%%
        \subsection{Information on \deltab}\label{Sect:info-deltab}
        
        In this section, we compare the information on \deltab brought by the survey, either raw information with $I(\deltab)$ defined by Eq.~\ref{Eq:def-Idb} or marginalised information with $I-\delta I$ defined by Eq.~\ref{Eq:def-dI}, to the prior information given by the $S$ matrix for the considered redshift bins.
        
\begin{figure}[!ht]
                \begin{center}
                        \includegraphics[width=.9\linewidth]{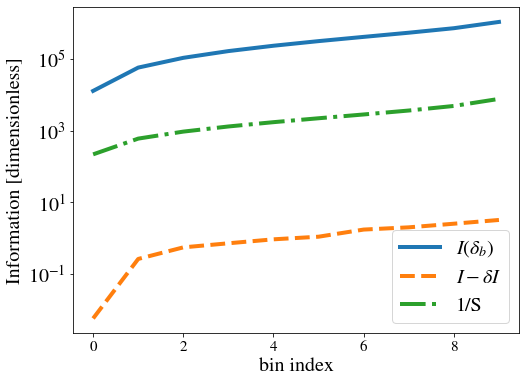}
                        \includegraphics[width=.9\linewidth]{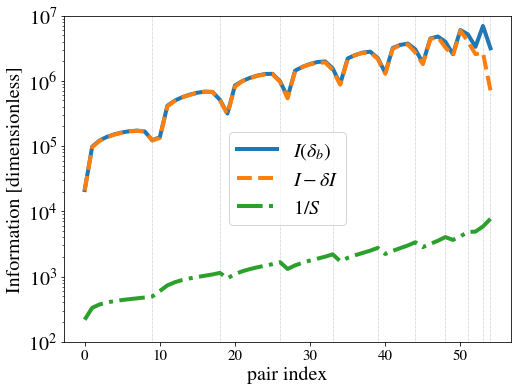}
                        \caption{Information on \deltab: Raw ($I$), marginalised ($I-\delta I$), and prior ($1/S$). \textit{Top:} Information in each bin of redshift for WL analyses of each bin separately. \textit{Bottom:} Information in each pair of redshift bins for a joint tomographic WL analysis. The 55 pairs are ordered as (0,0)(0,1)...(0,9)(1,1)..(1,9)...(8,8)(8,9)(9,9). The light vertical lines indicate the jump of the first redshift bin, i.e., the leftmost line is at (0,9) before the jump to (1,1).
                        }
                        \label{Fig:I-dI-S-WL}
                \end{center}
        \end{figure}

        We first consider the WL-only analysis. The first case of interest is to try to constrain \deltab in a redshift bin by only using the information in this bin. The result is shown in the top panel of Fig.~\ref{Fig:I-dI-S-WL}, where we plot $I(\deltab)$, $I-\delta I$, and $1/S$ in each of the ten redshift bins. The raw information (solid blue) is an order of magnitude larger than the prior (dash-dotted green), which appears to mean that the data can constrain \deltab very well. However, the marginalised information (dashed orange) falls by $\sim$3 orders of magnitude compared to the unmarginalised information, making it lower than the prior by $\sim$2 orders of magnitude. This comes from a high cancellation of $I(\deltab)$ and $\delta I$ due to high degeneracy between \deltab and cosmological parameters. We found that this is, in fact, not peculiar to \deltab; with a single redshift bin, most cosmological parameters are near perfectly degenerate, and the Fisher matrix has very poor conditioning.
        
        This fact motivated us to pursue the single bin case no longer and instead analyse the tomographic case of a joint analysis of all bins and further pairs of bins together. The result is shown in the bottom panel of Fig.~\ref{Fig:I-dI-S-WL}, where we plot $I(\deltab)$, $I-\delta I$, and $1/S$ in each of the 55 redshift pairs. Explicitly, each of these quantities is now a (55,55) matrix, and the figure shows the diagonal of these matrices.
        
        In this tomographic case, the raw information is larger than the prior by an even larger amount, $\sim$2 orders of magnitude, than in the single bin case. The parameter degeneracies are now all lifted, both for cosmological parameters and for nuisance parameters such as \deltab. We attribute this to the fact that the tomography provides information on the history of expansion and growth. Hence, here the marginalised information $I-\delta I$ sits almost perfectly on top of the unmarginalised information $I(\deltab)$, with only slight loss in the high-redshift pairs.
        
\begin{figure}[!ht]
                \begin{center}
                        \includegraphics[width=.9\linewidth]{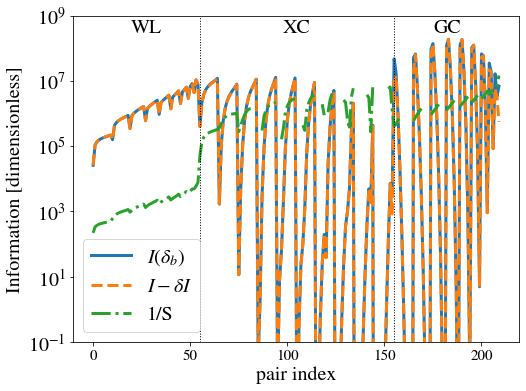}
                        \caption{Information on \deltab in each redshift bin pair in a tomographic 3x2pt analysis (WL + GC + XC). There are 210 pairs: 55 for WL, 100 for XC, and 55 for GC. The pairs are ordered as in Fig.~\ref{Fig:I-dI-S-WL}. The only difference is that XC has more pairs because there is no symmetry $(i,j)\leftrightarrow(j,i)$.}
                        \label{Fig:I-dI-S-3x2-allbins}
                \end{center}
        \end{figure}

        Next, we studied the 3x2pt analysis, which combines the WL auto-spectra, the GC auto-spectra, and the XC cross-spectra. There are now 210 bin pairs: 55 for WL, 55 for GC, and 100 for XC because the latter is not symmetric. Figure~\ref{Fig:I-dI-S-3x2-allbins} shows $I(\deltab)$, $I-\delta I,$ and $1/S$ in each of the 210 redshift pairs. We chose to order the pairs with WL first, then XC, and then GC; the corresponding intervals are indicated by vertical dashed lines in the figure.
        
        The marginalised information is oftentimes indistinguishable from the unmarginalised one and is larger than the prior. Deviations from this behaviour are visible as breaks of the lines in the XC and GC pairs. We cut the y-axis of the plot to zoom in on the interesting region. Otherwise, the blue and orange lines fall extremely low, making the plot unreadable. For these pairs, the two redshift kernels overlap extremely little, so that the signal is practically absent. This does not happen for WL because the WL kernels are broad in redshift, integrating everything between the source and the observer so that all kernels overlap strongly. However, the GC kernels are much more localised, and there is, for instance, practically no overlap between the first and last redshift bins. The cross-correlation XC will also vanish when measured between a low or median WL bin and a high GC bin. We note that the prior (green line) also breaks in these pair cases because \texttt{PySSC} sets the $S$ matrix element to 0 when the overlap of kernels is too small for a reliable computation. In practice, however, this is not an issue because in these pairs, the signal is too low to contain any cosmological information. Essentially, Fig.~\ref{Fig:I-dI-S-3x2-allbins} shows that \deltab can be well constrained for all relevant portions of the data.
        
        A theoretical concern that could come to mind, already for the tomographic WL analysis and even more for the 3x2pt analysis, is the physical significance of these numerous \deltab parameters: one for each bin pair. Through Eq.~\ref{Eq:def-U}, they change only the observable for that bin pair; physically, they are the survey matter density contrast averaged over redshift with the product of the two kernels of the pair. The concern is that all these pairs probe a similar redshift range, and that therefore the parameters should be strongly correlated. The answer is that this is indeed the case, and in practice, the correlation of these parameters is quantified by the $S$ matrix, which has high off-diagonal correlation coefficients. When the formulae developed in Sect.~\ref{Sect:precision-matrix} for the likelihood or Fisher forecast are used, the \deltab parameters do not appear in the computation, and all the formulae are well behaved numerically. In a different context, for a different kind of analysis that explicitly has \deltab nuisance parameters, there could be numerical problems with their number and high correlation. In that case, Appendix~\ref{App:ext-Sij} devises an analysis with fewer less strongly correlated \deltab parameters, at the expense of needing more observable predictions.
        
        %%%%%%%%%%%%%%%%%%%%
        \subsection{Asymptotic errors}\label{Sect:asymto-errors}
        
        In this section, we determine the change in the Fisher constraints when an infinitely thin or wide prior $\mathcal{N}(0,S)$ is assumed. In the single bin case, this corresponds to the respective cases $S\rightarrow 0$ and $S\rightarrow \infty$.
        
        The case of an infinitely thin prior is simple. From Eq.~\ref{Eq:Fisher-singlebin} (single bin case) or Eq.~\ref{Eq:Fisher-multibins} (multi-bin case), the Fisher constraint asymptote to the noSSC one. This happens because the prior fixes \deltab to 0 in each bin, so that there is no SSC effect.
        
        The case of an infinitely wide prior is more interesting. It corresponds to \deltab being completely free, $S\rightarrow \infty$. It might be motivated by the observation of Sect.~\ref{Sect:info-deltab} that the data appear to constrain \deltab much better than the prior, so that the latter seems not to be needed.
        
        The Fisher matrix was already given in Eq.~\ref{Eq:Fisherlike-freedb}, but for the parameter vector $\theta'=(\theta,\deltab)$. After marginalisation over $\deltab$, the inverse Fisher is
        \ba
        F^{-1} &= F^{-1}_\mr{noSSC} + Z^T \ \left(I(\deltab)-\delta I\right)^{-1} \ Z.
        \ea
        The corresponding marginalised error bars can be compared to the fiducial case in which $S$ is given by the theoretical prediction from \texttt{PySSC}. The ratios of these two cases for cosmological parameters (after marginalisation over all the others, as well as the nuisance parameters) are given in Table~\ref{Tab:increase-err-noprior}. This conservative no-prior analysis would increase the cosmological errors by unacceptably large factors between 1.1 ($\Omega_b$, WL-only) and 5 ($w_a$, WL-only), with an average of 2.8 for WL-only and 2.5 for 3x2pt.
        
        \begin{table}   
                \begin{center}
                        \begin{tabular}{c|c|c|c|c|c|c|c}
                                & $\Omega_m$ & $\Omega_b$ & $w_0$ & $w_a$ & $h$ & $n_S$ & $\sigma_8$ \\
                                \hline
                                WL & 3.12 & 1.12 & 3.87 & 5.00 & 1.80 & 2.11 & 2.55 \\
                                3x2 & 3.49 & 1.27 & 3.13 & 3.56 & 1.71 & 1.41 & 2.96
                        \end{tabular}
                \end{center}
                \caption{Ratio of the error bar for each cosmological parameter between the case of no prior on \deltab and the case of the fiducial prior. Second line: Case of a tomographic WL analysis. Third line: Case of a 3x2pt analysis.}
                \label{Tab:increase-err-noprior}
        \end{table}
        
        At first, this may seem at odds with the result of Sect.~\ref{Sect:info-deltab} that the data appear to constrain \deltab much better than the prior. However, this happens because the data mostly constrain each \deltab independently, while, as discussed at the end of Sect.~\ref{Sect:info-deltab}, the prior has strong off-diagonal correlations, which encode the fact that the \deltab probe similar redshift ranges. In the no-prior case, we therefore assign much freedom to the data, which blows up the constraints.
        
        %%%%%%%%%%%%%%%%%%%%
        \subsection{Requirements for reaching robust cosmic errors}\label{Sect:requirements}
        
        The results of section~\ref{Sect:asymto-errors} pose the more general question of how precisely we need to model the super-sample covariance in order for cosmological constraints to be robust. This question is better posed at the covariance level than at the likelihood level used in Sect.~\ref{Sect:asymto-errors}. We focus on the impact on the error bars, although, as noted by \cite{Sellentin2019} using an incorrect covariance matrix is also equivalent to analysing a biased data set.
        
        To answer this question, we let the SSC responses vary both in amplitude and scale dependence and forecast the variation in cosmological constraints with respect to the case with the fiducial responses.
        
        %%%%%%%%%%
        \subsubsection{Weak-lensing-only case}\label{Sect:reqs-WL}
        
        For the WL case, we introduced two parameters, $A_0$ and $A_1$, which change the response as
        \ba\label{Eq:changing-responses-model1}
        \partial_{\deltab} C_\ell \rightarrow A_0 \left(1+ A_1 \frac{\ell-\ell_0}{2\ell_0}\right) \times \partial_{\deltab} C_\ell,
        \ea
        with $\ell_0$ a pivot scale chosen to be the centre of the multipole range so that the fraction varies by 1 between $\ell_\mr{min} \sim 0$ and $\ell_\mr{max} = 2\ell_0$.\\
        Based on the developments of Sect.~\ref{Sect:update-applications}, we can compute a Fisher forecast without building and inverting a new covariance matrix for each value of $(A_0,A_1)$ we wish to test. Instead, we first computed the Fisher forecast for the Gaussian covariance matrix and then used Eq.~\ref{Eq:iFisher-multibins} to determine the inverse Fisher matrix. We found that the bottleneck is the computation of $I(\deltab)$, whose brute-force computation takes $\sim3$s on a laptop with i7 CPU. By comparison, a covariance inversion takes $\sim120$s on the same laptop, which would make the forecast painful to run when many values for $A_0$ and $A_1$ need to be tested.
        This speedup was achieved without using the fact that the Gaussian covariance is diagonal in terms of multipoles.
        
\begin{figure}[!ht]
                \begin{center}
                        \includegraphics[width=.9\linewidth]{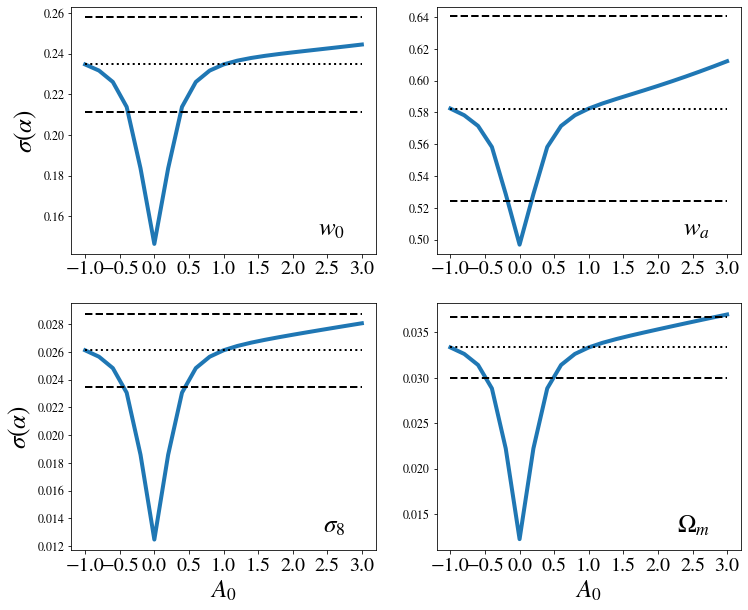}
                        \includegraphics[width=.9\linewidth]{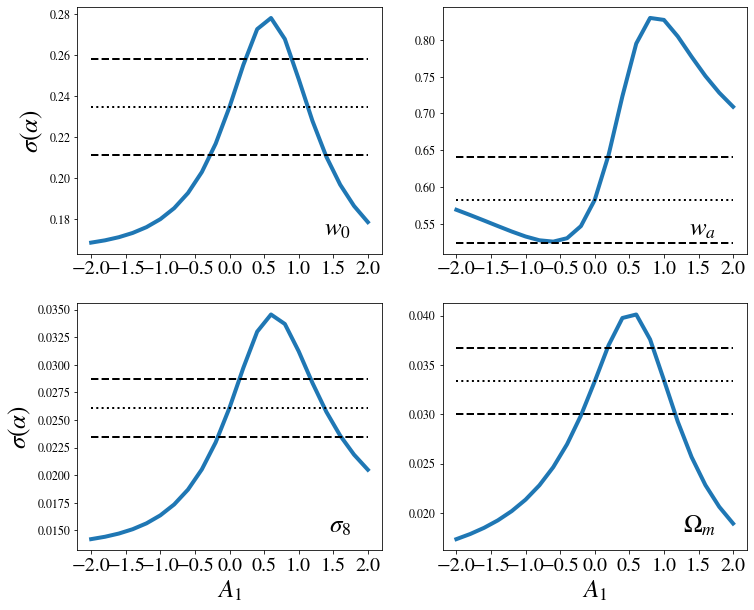}
                        \caption{Forecasted errors on the dark energy equation of state, $\sigma_8$ and $\Omega_m$ for a tomographic WL analysis. \textit{Top two rows:} Constraints as a function of $A_0$ (which changes the amplitude of the SSC response). \textit{Bottom two rows:} Constraints as a function of $A_1$ (which changes the scale dependence of the SSC response). Dotted lines show the value for the fiducial covariance, $(A_0,A_1)=(1,0)$, and the dashed lines show $\pm$ 10\% around the dotted line.}
                        \label{Fig:err-vs-A_0-A_1-WL-allbins}
                \end{center}
        \end{figure}

        We can then study how the marginalised errors on cosmological parameters change with $A_0$ and $A_1$. The results are shown in Fig.~\ref{Fig:err-vs-A_0-A_1-WL-allbins} for the most affected parameters, which are in fact the usual targets of WL experiments: the dark energy equation of state, $\sigma_8$ and $\Omega_m$.
        
        We let $A_0$ run between -1 and 3, that is, between predicting the response with a sign mistake and misestimating the response by a factor of 3. We let $A_1$ run between -2 and 2, which is a scale dependence that is strong enough to cause the response to vanish at the minimum or maximum bound of the multipole range. The plots also contain dashed horizontal lines to indicate when the errors deviate from the fiducial case by more than 10\%, in line with the requirements of stage IV missions \cite[e.g. Euclid,][]{Euclid-redbook}
        
        The results show that the branch in which $A_0$ is negative is symmetric to the positive branch. This happens because $A_0$ is squared in the covariance equation,
        %which can be traced back to the base SSC equation Eq.~\ref{Eq:SSC-base-eq} where the response appears twice,
        so that an overall sign mistake has no impact. As $A_0$ increases (in absolute value), so do the cosmological errors. The requirement that errors should have a precision of 10\% constrains the allowed value of $A_0$. The strongest constraints come from $\Omega_m$, giving $A_0 \in [0.47,2.82]$. We note that this is a quite large interval: response predictions can be under- or over-estimated by a factor larger than 2.
        
        For $A_1$ the curves are no longer symmetric and have a steep slope near the fiducial case $A_1=0,$ so that they easily exceed the 10\% requirements. Therefore, satisfying the required precision places strong constraints on $A_1$; the strongest constraint comes from $\sigma_8$, giving $A_1 \in [-0.16,0.15]$.
        
        In summary, WL constraints are rather robust to a mismodelling of the amplitude of its response. However, the constraints are more sensitive to the scale dependence of the response, which must be modelled at the 15\% level.
        
        %%%%%%%%%%
        \subsubsection{3x2pt case}\label{Sect:reqs-3x2}
        
\begin{figure}[!ht]
                \begin{center}
                        \includegraphics[width=.9\linewidth]{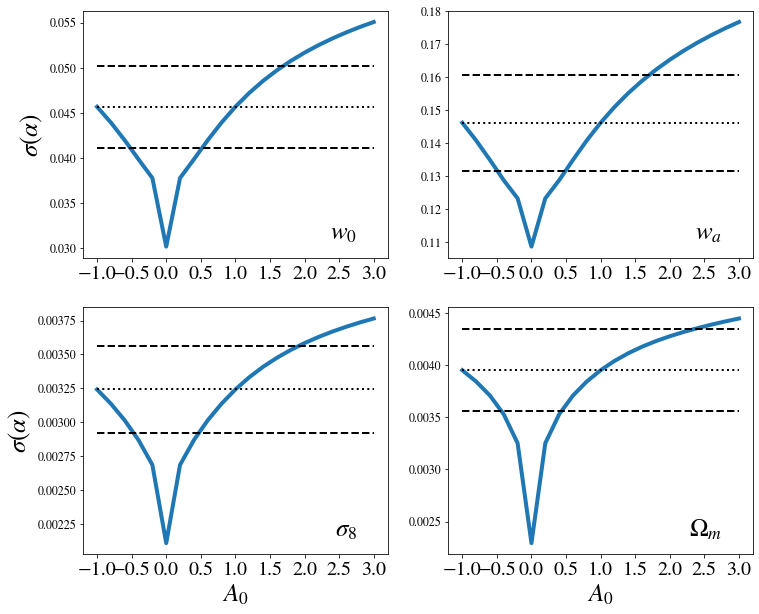}
                        \includegraphics[width=.9\linewidth]{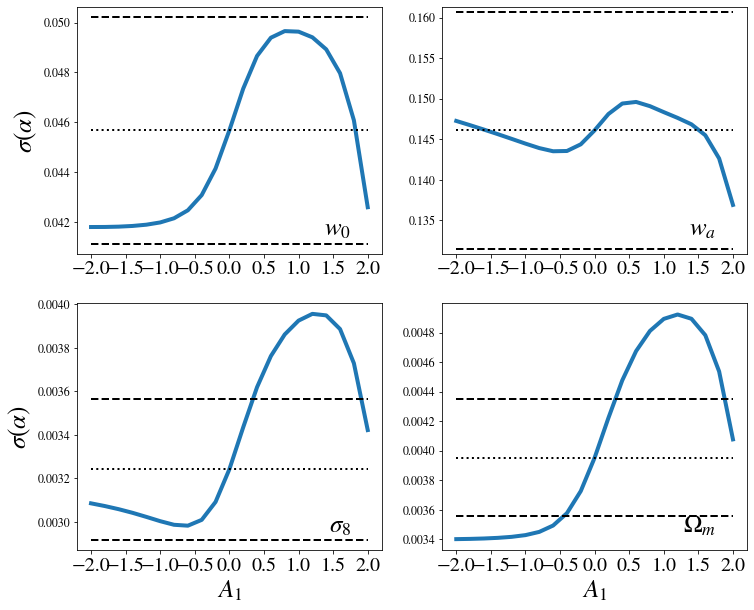}
                        \caption{Forecasted errors on the dark energy equation of states, $\sigma_8$, and $\Omega_m$ for a 3x2pt analysis. \textit{Top two rows:} Constraints as a function of $A_0$, which changes the amplitude of the SSC response. \textit{Bottom two rows:} Constraints as a function of $A_1$, which changes the scale dependence of the SSC response. Dotted lines show the value for the fiducial covariance, and the dashed lines are $\pm$ 10\% around the dotted line.}
                        \label{Fig:err-vs-A_0-A_1-3x2-allbins}
                \end{center}
        \end{figure}

        For the 3x2pt case, that is, when WL is combined with GC, including XC, we can first use the same parametrisation Eq.~\ref{Eq:changing-responses-model1}, with $A_0$ and $A_1$ being the same for all three probes. We can then observe the impact of varying these parameters, which change the amplitude and scale dependence of all probes together. The results are shown in Fig.~\ref{Fig:err-vs-A_0-A_1-3x2-allbins} for the parameters that are usually most affected: dark energy, $\sigma_8,$ and $\Omega_m$.
        
        The forecasts are generally more robust to these mismodelling parameters. For $A_0$, the stringent constraints are set by $w_0$, which requires $A_0 \in [0.53,1.70]$. This is slightly stricter than the WL case, but still quite large. For $A_1$, the stringent constraints come from $\Omega_m$ , which requires $A_1 \in [-0.44,0.29]$. The scale dependence must therefore be modelled at $\sim$30\% precision; this constraint is twice looser than for the WL-only case.
        
\begin{figure}[!ht]
                \begin{center}
                        \includegraphics[width=.9\linewidth]{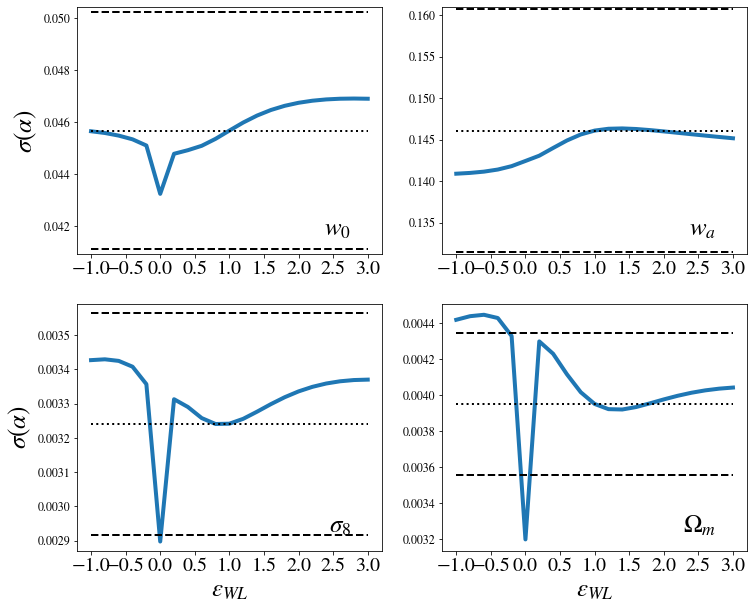}
                        \includegraphics[width=.9\linewidth]{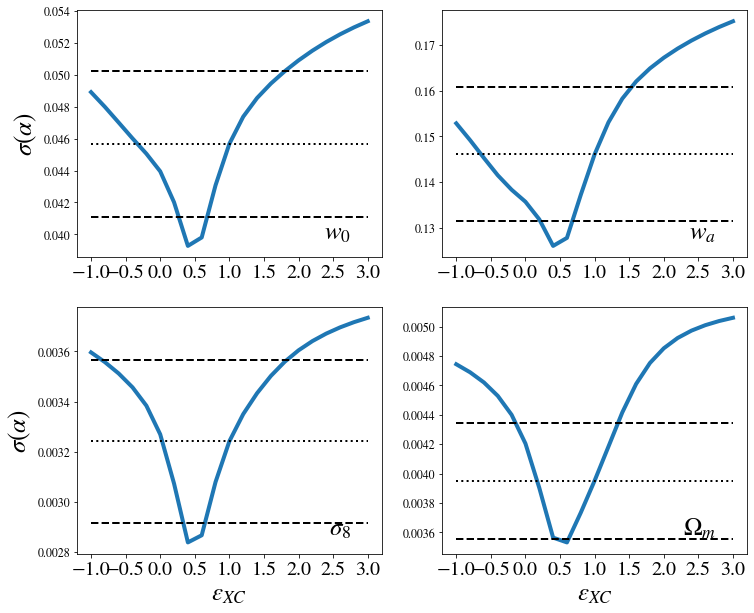}
                        \includegraphics[width=.9\linewidth]{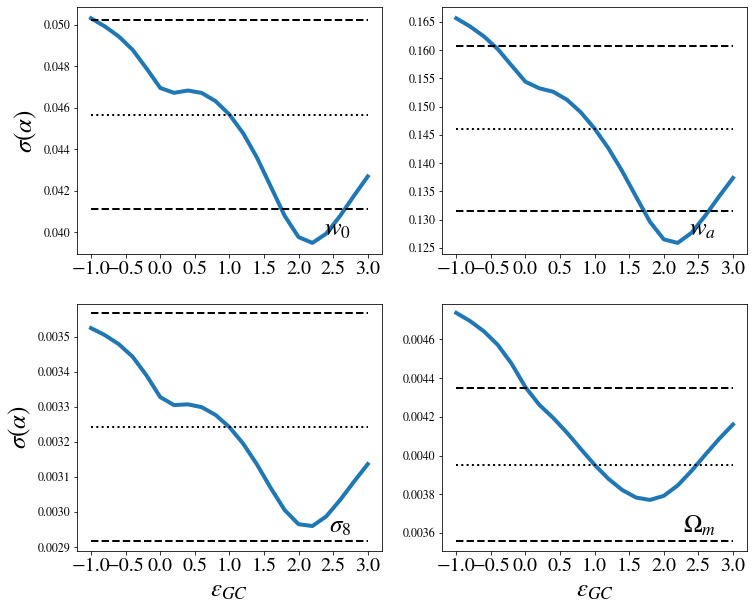}
                        \caption{For a 3x2pt analysis, forecast errors on dark energy, $\sigma_8,$ and $\Omega_m$.
                                \textit{Top:} As a function of $\epsilon_{WL}$, which changes the WL response alone.
                                \textit{Centre:} As a function of $\epsilon_{XC}$, which changes the cross-correlation response alone.
                                \textit{Bottom:} As a function of $\epsilon_{GC}$, which changes the GC response alone. Dotted line: Value for the fiducial covariance. Dashed lines: $\pm$ 10\% around the dotted line.}
                        \label{Fig:err-vs-e_probes-3x2-allbins}
                \end{center}
        \end{figure}

        These results hold for a common mismodelling of all responses of all probes, for example in a separate universe approach, if we have a systematic common to all probes in the simulations, or in a theoretical approach, if we were missing a term in our equations.
        To study the case when the physics of a single probe is modelled incorrectly, we defined a new model for the responses with new parameters $\epsilon_p$ for the amplitude of the response of each probe independently,
        \ba\label{Eq:changing-responses-model2}
        \partial_{\deltab} C_\ell^p \rightarrow \epsilon_p \left(1+ A_1 \frac{\ell-\ell_0}{2\ell_0}\right)  \times \partial_{\deltab} C_\ell^p,
        \ea
        where $(\epsilon_{WL},\epsilon_{GC},\epsilon_{XC})$ modulate the response of the power spectrum of WL, GC, and XC respectively. Fig.~\ref{Fig:err-vs-e_probes-3x2-allbins} shows the impact of these parameters on the parameters that are usually most affected (dark energy equation of state parameters, $\sigma_8,$ and $\Omega_m$), 
        
         $\epsilon_{WL}$ has little impact on the constraints, which easily stay within the requirements for most of the explored range of values. Only $\Omega_m$ discards some values. In particular, the constraints on dark energy are very robust throughout all explored ranges. The error bars are also quite robust to $\epsilon_{GC}$ , although to a lesser extent than for $\epsilon_{WL}$. For instance, constraints on $\sigma_8$ stay within the requirements in all explored ranges. Lower and upper values on $\epsilon_{GC}$ are set by different cosmological parameters: to meet the 10\% requirement, $\Omega_m$ requires $\epsilon_{GC} \geq 0.01$ and $w_a$ requires $\epsilon_{GC} \leq 1.71$. The combined allowed range is thus $\epsilon_{GC} \in [0.01,1.71]$ , which is quite large. By contrast, in the case of $\epsilon_{XC}$ , the errors exceed the requirements much more easily. The lower constraint is $\epsilon_{XC} \geq 0.69$ due to $w_0$, and the upper constraint is $\epsilon_{XC} \leq 1.34$ due to $\Omega_m$. The lower constraint could be relaxed by adopting a slightly looser requirement, as the curves drop just below the -10\% line, although we do not enter into these considerations. In summary, the amplitude of this response must be modelled at a precision of $\sim$30\%.
        
        It is expected that variations in the XC response have the most important effect because it has been shown that this cross correlation is of the utmost importance to control systematics and unlock the full power of the 3x2pt analysis \citep{Tutusaus2020}.
        
        In summary, 3x2pt constraints are more robust to SSC mismodelling than WL-only constraints. The most important quantities to model correctly are the amplitude of the XC response and the scale dependence of all responses. Both need to be calibrated to the $\sim$30\% level.
        
        %%%%%%%%%%%%%%%%%%%%%%%%%%%%%%%%%%%%%%%%%%%%%%%%%%%%%%%%%%%%%%%%%%%
        
        \section{Conclusion}\label{Sect:conclusion}
        
        Within the framework of the Sij approximation to SSC, we have presented a new formula in which SSC is an update to the precision (inverse covariance) matrix. This update can be computed numerically in a time comparable to that of the noSSC precision matrix and can take advantage of any sparsity in the noSSC covariance. We showed practical examples for two types of cosmological probes: number counts, and power spectrum. We presented the application of this formula to the computation of the likelihood or a Fisher forecast as an update to the noSSC case. The Fisher matrix (and inverse Fisher matrix) were then used to prove the formal equivalence of three approaches to super-sample covariance: (i) classically at the covariance level, (ii) at the likelihood level through dedicated nuisance parameters, and (iii) with a QE to estimate the super-sample modes. Examination of these three approaches allowed us to give a physical interpretation of the various quantities entering the formula. Although these approaches are formally equivalent, we argue that the second approach, at the likelihood level, is conceptually more interesting for two reasons: (i) the \deltab nuisance parameters can be examined to track systematics, which would cause them to deviate significantly from their prior (e.g. a systematic inducing an excess of power in some bins would push \deltab>0 for these bins), and (ii) it provides a natural framework to determine whether the data prefer low density (\deltab<0) at low redshift. This potential explanation has been put forward in the past for acceleration or dark energy or the $H_0$ tension.
        
        Furthermore, we presented a numerical application of this framework to a mock stage IV weak-lensing survey. Namely, we considered a WL-only and a WL + GC + XC (also known as 3x2pt) case. We showed that although the \deltab parameters are well constrained, we cannot recommend to remove their theoretical prior  because it would strongly degrade the cosmological constraints. We also examined the change induced by mismodelling on the cosmological errors, and especially how much mismodelling can be allowed to still satisfy the requirement of a precision of 10\%  for stage IV surveys. This is of interest as there are always theoretical or simulation uncertainties in the prediction of these responses. We found that large mispredictions of the response amplitude are allowed for the WL-only case, but the scale dependence must be predicted more precisely at the $\sim15\%$ level. For the 3x2pt case, large amplitude mispredictions are possible if it is common to all probes, but a misprediction specific to the cross-spectrum signal (XC=GC-WL power spectrum) is more problematic. This amplitude needs to be calibrated to a precision of $\sim$30\%. The scale dependence of the response must also be predicted precisely, at the $\sim30\%$ level, although the requirement is less strict than for the WL-only case. In practice, this is the target of convergence that should be aimed for in the comparison of the response models or simulations.
        
        %%%%%%%%%%%%%%%%%%%%%%%%%%%%%%%%%%%%%%%%%%%%%%%%%%%%%%%%%%%%%%%%%%%%
        \section*{Acknowledgements}
        \vspace{0.2cm}
        
        We thank St\'{e}phane Ili\'{c} for fruitful exchanges and Joachim Harnois-D\'{e}raps for proofreading the manuscript.\\
        F.L. acknowledges support from the Swiss National Science Foundation through grant number IZSEZ0\_207059, and from Progetto di Eccellenza 2022 of the Physics and Astronomy department at the University of Padova "Physics of the Universe", within the grant "CosmoGraLSS — Cosmology with Gravitational waves and Large Scale Structure". JC and LL acknowledge support from a SNSF Eccellenza Professorial Fellowship (No. 186879). AMD acknowledges support from Tommala Foundation for research in gravity as well as Boninchi Foundation. SGB was supported by CNES, and focused on the \textit{Euclid} mission. The project leading to this publication has received funding from Excellence Initiative of Aix-Marseille University -A*MIDEX, a French "Investissements d'Avenir" program (AMX-19-IET-008 -IPhU). AG's work is supported by the McGill Astrophysics Fellowship funded by the Trottier Chair in Astrophysics, as well as the Canadian Institute for Advanced Research (CIFAR) Azrieli Global Scholars program and the Canada 150 Programme. \\
        The calculations in this article made use of \texttt{\href{https://class-code.net}{class}} \citep{Blas2011}, \href{https://cosmosis.readthedocs.io/en/latest/}{\texttt{cosmosis}} by \cite{Zuntz2015}, \texttt{astropy} \citep{astropy,astropy2}, \texttt{matplotlib} \citep{Hunter2007}, \texttt{scipy} \citep{scipy}, and \texttt{numpy} \citep{numpy}. 
        
        %%%%%%%%%%%%%%%%%%%%%%%%%%%%%%%%%%%%%%%%%%%%%%%%%%%%%%%%%%%%%%%%%%%%
        \bibliographystyle{aa}
        \bibliography{bibliography}
        %%%%%%%%%%%%%%%%%%%%%%%%%%%%%%%%%%%%%%%%%%%%%%%%%%%%%%%%%%%%%%%%%%%%
        
        \begin{appendix}
        
        %%%%%%%%%%%%%%%%%%%%%%%%%%%%%%%%%%%%%%%%%%%%%%%%%%%%%%%%%%%%%%%%%%%
        \section{Quadratic estimator for \deltab}\label{App:QE}
        
        We wish to estimate the background density over the full sky, $\delta_b(z) = \int \frac{\dd^2 \hn}{4\pi} \ \delta_m(r\hn,z) $, which is proportional to the monopole of the matter density field, $\delta_b(z) = \left. a_{00}(\delta_m) \middle/ \sqrt{4\pi}\right.$.
        
        We use notations from the seminal article of \cite{Okamoto2003} with slight variations to avoid problems at $L=0$, and consider a single pair of fields. The estimator for the mode $(L,M)$ has the general form
        \ba
        \nonumber d^M_L =& \ A_L \sum_{\ell_1,m_1} \sum_{\ell_2,m_2} (-1)^M \threeJm{\ell_1}{\ell_2}{L}{m_1}{m_2}{-M} \\
        & \times g_{\ell_1,\ell_2}(L) \ a_{\ell_1,m_1} \ a_{\ell_2,m_2} .
        \ea
        We apply this to $L=M=0$, which leads to an important collapse. The Wigner 3J symbol enforces the triangular conditions  $|\ell_1-\ell_2|\leq L=0$ and $m_1+m_2-M=0$, so that we have $\ell_1=\ell_2$ and $m_2=-m_1$. We obtain
        \be
        d^0_0 = A_0 \sum_{\ell,m} g_{\ell,\ell}(0) \ a_{\ell,m} \ a_{\ell,m}^*,
        \ee
        where we absorbed multiplicative factors into $g_{\ell,\ell}(0)$ and used $a_{\ell,-m}= (-1)^m a_{\ell,m}^*$. Then we use the definition of the power spectrum estimator in full sky $\hat{C}_\ell = \frac{1}{2\ell+1} \sum_m a_{\ell,m} a_{\ell,m}^*$, absorb the multiplicative factors again, and obtain        \be\label{Eq:d00}
        d^0_0 = A_0 \sum_{\ell} g_{\ell,\ell}(0) \ \hat{C}_\ell.
        \ee
        The normalisation $A_0$ has to be chosen such that the estimator is unbiased.
        In our case, $<\hat{C}_\ell>_{\delta_b\, \mr{fixed}} = \overline{C}_\ell(\theta) + \deltab \ \partial_{\deltab} C_\ell$, so
        \ba
        <d^0_0>= A_0 \sum_{\ell} g_{\ell,\ell}(0) \overline{C}_\ell(\theta) + \deltab \ A_0 \sum_{\ell} g_{\ell,\ell}(0) \partial_{\deltab} C_\ell.
        \ea
        Our estimator therefore is
        \ba
        \hat{\deltab} &= d^0_0 - A_0 \sum_{\ell} g_{\ell,\ell}(0) \overline{C}_\ell(\theta) \\
        &= A_0 \sum_{\ell} g_{\ell,\ell}(0) \left(\hat{C}_\ell-\overline{C}_\ell(\theta)\right), \label{Eq:hdb-funct-A0gll0}
        \ea
        where the normalisation follows
        \be\label{Eq:A-funct-gll0}
        A_0 = \left(\sum_{\ell} g_{\ell,\ell}(0) \ \partial_{\deltab} C_\ell\right)^{-1}.
        \ee
        We are left with optimising the weight $g_{\ell,\ell}(0)$ (noting that a constant multiplicative factor in its definition is unimportant as it would cancel out in the ratio in Eq.~\ref{Eq:hdb-funct-A0gll0}). This optimisation is performed by minimising the variance of the estimator,
        \be\label{Eq:var-d00}
        \mathrm{Var}(d^0_0) = \lbra d^0_0 \, d^0_0 \rbra - \lbra d^0_0\rbra^2 = A_0^2 \sum_{\ell,\ell'} g_{\ell,\ell}(0) g_{\ell',\ell'}(0) \mathcal{C}_{\ell,\ell'},
        \ee
        where $\mathcal{C}_{\ell,\ell'}$ is the covariance of $\hat{C}_\ell$. A significant detail here is that all the brackets mean average at \deltab fixed, 
        \ba
        \mathcal{C}_{\ell,\ell'} = <\hat{C}_\ell \ \hat{C}_{\ell'}>_{\delta_b\, \mr{fixed}} - <\hat{C}_\ell >_{\delta_b\, \mr{fixed}} <\hat{C}_{\ell'}>_{\delta_b\, \mr{fixed}}.
        \ea
        Hence $\mathcal{C}_{\ell,\ell'}$ is the noSSC part of the covariance.
        
        After some algebra, we find that the minimisation yields
        \ba
        \sum_{\ell'} \mathcal{C}_{\mr{noSSC};\ell,\ell'} \ g_{\ell',\ell'}(0) = \partial_{\deltab} C_\ell \times \mr{cst}.
        \ea
        We can therefore take
        \be\label{Eq:QE_gll0}
        g_{\ell,\ell}(0) = \sum_{\ell'} \mathcal{C}^{-1}_{\mr{noSSC};\ell,\ell'} \ \partial_{\deltab} C_{\ell'}.
        \ee
        This also gives the normalisation of the estimator,
        \be\label{Eq:QE_A0}
        A_0 = \left(\sum_{\ell,\ell'} \partial_{\deltab} C_\ell \  \mathcal{C}^{-1}_{\mr{noSSC};\ell,\ell'} \ \partial_{\deltab} C_{\ell'} \right)^{-1}.
        \ee
        
        In summary, the quadratic estimator takes the form
        \ba
        \hat{\deltab} = \frac{1}{N} \sum_{\ell} g_{\ell,\ell}(0) \, (\hat{C}_\ell-\overline{C}_\ell(\theta)),
        \ea
        with $N = A_0^{-1}$ given by Eq.~\ref{Eq:QE_A0} and $g_{\ell,\ell}(0)$ given by Eq.~\ref{Eq:QE_gll0}
        
        %%%%%%%%%%%%%%%%%%%%%%%%%%%%%%%%%%%%%%%%%%%%%%%%%%%%%%%%%%%%%%%%%%%
        \section{Inverse Fisher matrix in the likelihood approach}\label{App:iFisher-lik}
        
        In the likelihood approach, the total Fisher matrix $F'=F(\mr{free \ \deltab})+F(\mr{prior})$ takes the form
        \ba
        F' = \begin{pmatrix} F^\mr{noSSC} & Y^T \\ Y & I(\deltab)\end{pmatrix} + \begin{pmatrix} 0 & 0 \\ 0 & \frac{1}{S} \end{pmatrix}.
        \ea
        For readability, we now write $I\equiv I(\deltab)$. In order to obtain the forecast constraints on $\theta$ marginalised over $\deltab$, we now need to invert $F'$ and extract the upper left $(n_\theta,n_\theta)$ block. For this, we shuffle terms around so that we can write $F'$ in the form $F'=F'_{(1)}+F'_{(2)}$ with
        \ba
        F'_{(1)} = \begin{pmatrix} F^\mr{noSSC} & 0 \\ 0 & \frac{1}{S}+I\end{pmatrix},
        \qquad \qquad
        F'_{(2)} = \begin{pmatrix} 0 & Y^T \\ Y & 0 \end{pmatrix}.
        \ea
        This makes $F'_{(1)}$ easy to invert as it is block-diagonal. For the second term, we remark that
        \ba
        F'_{(2)} = \begin{pmatrix} Y & 0 \\ 0 & 1\end{pmatrix} \begin{pmatrix} 0 & 1 \\ 1 & 0 \end{pmatrix} \begin{pmatrix} Y^T & 0 \\ 0 & 1 \end{pmatrix},
        \ea
        which puts $F'$ in a form amenable to using the Woodbury identity one more time. After some algebra, we find
        \ba
        F'^{-1} &= F'^{-1}_{(1)} + \frac{S}{1 + (I-\delta I) S} \begin{pmatrix} Z^T Z & Z^T \\ Z & \frac{\delta I \, S}{1+I S}. \end{pmatrix}
        \ea
        Shuffling terms around, we obtain the cleaner form
        \ba
        F'^{-1} &= \begin{pmatrix} F^{-1}_{\mr{noSSC}} & 0 \\ 0 & 0\end{pmatrix} + \frac{S}{1 + (I-\delta I) S} \begin{pmatrix} Z^T Z & Z^T \\ Z & 1\end{pmatrix}.
        \ea
        
        %%%%%%%%%%%%%%%%%%%%%%%%%%%%%%%%%%%%%%%%%%%%%%%%%%%%%%%%%%%%%%%%%%%
        \section{Extension of the Sij approximation to probe-independent $\delta_b$s}\label{App:ext-Sij}
        
        A LSS observable $\mathcal{O}(i,\alpha)$ can generally be cast in the form of a light-cone integral of some density $\mathfrak{o}(i,\alpha)$,
        \be
        \mathcal{O}(i,\alpha) = \int \dd V \ \mathfrak{o}(i,\alpha).
        \ee
        (We note that for correlation functions of order $n\geq 2$, this requires the use of Limber's approximation.)\\
        In this case, the equation for super-sample covariance is given by \citep[e.g.][]{Lacasa2016}
        \ba \label{Eq:exact-SSC}
        \nonumber \Cov_\mr{SSC}\left[\mathcal{O}(i,\alpha),\mathcal{O}(j,\beta)\right] =& \int \dd V_{12} \, \frac{\partial \mathfrak{o}(i,\alpha)}{\partial\delta_b}(z_1) \, \frac{\partial \mathfrak{o}(j,\beta)}{\partial \delta_b}(z_2) \\ 
        & \times \sigma^2(z_1,z_2),
        \ea
        where $\sigma^2(z_1,z_2)=\Cov\left[\deltab(z_1),\deltab(z_2)\right]$ is the covariance of the background shifts of infinitesimal redshift slices. Starting from this equation, \cite{Lacasa2019} derived the general Sij approximation Eq.~\ref{Eq:SSC-base-eq} by assuming that the responses vary slowly with redshift compared to $\sigma^2(z_1,z_2)$. Now this approximation can be refined by cutting the integrals into small pieces and performing the approximation in each piece. Concretely, we start by defining a series of functions $\Phi_n(z)$ that sum to unity over a redshift range suitable for the integrals for all considered observables,
        \be
        \sum_{n=1}^{n_\Phi} \Phi_n(z) = 1 \qquad \forall z \in [z_\mr{min},z_\mr{max}].
        \ee
        An example of such $\Phi_n$ is series of non-intersecting top-hat functions that cover $[z_\mr{min},z_\mr{max}]$. Gaussian-like functions are another possibility. They have the advantage of avoiding ringing effects in later integrals.\\
        The LSS observable then takes the form
        \be
        \mathcal{O}(i,\alpha) = \sum_n \mathcal{O}(i,\alpha)_n \ \ \mr{with} \ \ \mathcal{O}(i,\alpha)_n = \int \dd V \; \Phi_n(z) \; \mathfrak{o}(i,\alpha), 
        \ee
        and its super-sample covariance is
        \be
        \nonumber \Cov_\mr{SSC}\left[\mathcal{O}(i,\alpha),\mathcal{O}(j,\beta)\right] = \sum_{m,n} \Cov_\mr{SSC}\left[\mathcal{O}(i,\alpha)_m ,\mathcal{O}(j,\beta)_n\right],
        \ee
        where we can apply the Sij approximation to $\Cov_\mr{SSC}\left[\mathcal{O}(i,\alpha)_m ,\mathcal{O}(j,\beta)_n\right]$:
        \ba
        \nonumber \Cov_\mr{SSC}\left[\mathcal{O}(i,\alpha)_m ,\mathcal{O}(j,\beta)_n\right] =& \int \dd V_{12} \ \Phi_m(z_1) \, \Phi_n(z_2) \\
        \nonumber & \times \frac{\partial \mathfrak{o}(i,\alpha)}{\partial \deltab}(z_1) \, \frac{\partial \mathfrak{o}(j,\beta)}{\partial \delta_b}(z_2) \\
        & \times \sigma^2(z_1,z_2)\\
        & \approx \partial_{\deltab} \mathcal{O}(i,\alpha)_m \ \partial_{\deltab} \mathcal{O}(j,\beta)_n \ S_{m,n},
        \ea
        with
        \be
        S_{m,n} = \int \dd V_1 \, \dd V_2 \, \frac{\Phi_m(z_1)}{\Delta V_m} \, \frac{\Phi_n(z_2)}{\Delta V_n} \ \sigma^2(z_1,z_2),
        \ee
        and
        \be
        \Delta V_m = \int \dd V \ \Phi_m(z).
        \ee
        where we assumed that the $\Phi_n$ functions have sufficiently narrow support for probe-specific kernels to be considered constant. In practical tests, we found that a width $\Delta z \sim 0.1$ is small enough. The SSC matrix thus has the same form as Eq.~\ref{Eq:SSC-USU_form},
        \be
        \mathcal{C}_\mr{SSC} = U \, S \, U^T,
        \ee
        where now $U$ is a dense rectangular matrix with the shape $(n_o,n_\Phi)$,
        \be
        U_{i_o,n} = \partial_{\deltab} \mathcal{O}(i,\alpha)_n.
        \ee
        Thus all the machinery defined in the main text of the article can be used here, as well; Eq.~\ref{Eq:precision-tot} for the inverse covariance, Eq.~\ref{Eq:update-loglike} for the likelihood, and Eq.~\ref{Eq:Fisher-multibins} and Eq.~\ref{Eq:iFisher-multibins} for the Fisher and inverse Fisher matrices.
        
        The key difference is that the $S_{m,n}$ matrix is now independent of the considered physical probe; indeed, it is the covariance matrix of probe-independent $\deltab$ parameters,
        \be
        S_{m,n} = \Cov\left[\delta_{b,m},\delta_{b,n}\right] \ \ \mr{with} \ \ \delta_{b,n}=\int \dd V \ \Phi_n(z) \ \deltab(z)
        \ee
        In the likelihood approach to SSC, this $\delta_{b,n}$ is the vector of nuisance parameters over which is to be marginalised. The difference with the main text of the article is that this set of parameters is now independent of the physical probe and does not grow when probes are combined. The number of these parameters  is defined by the width of the $\Phi_n$ functions, and optimisation can be studied to keep this width as large as possible while maintaining the validity of the Sij approximation.
        
        According to the results of \cite{Lacasa2019}, a width $\Delta z=0.1$ is already small enough to maintain the validity of the approximation. For a stage IV galaxy survey out to $z_\mr{max}=2$, 20 parameters or even fewer are therefore enough to analyse all physical probes. This compares favourably to the 55 \deltab parameters of the WL-only case in the main text or to the 210 parameters of the 3x2pt case. Moreover, this number is small enough not to be a limiting factor in an MCMC analysis, especially in the context of a combined probe analysis where the number of other nuisance parameters is already high. The only downside to this approach is that the pipelines need to be modified to output all the $\partial_{\deltab} \mathcal{O}(i,\alpha)_m$, an output with a size 20 times the size of the observable prediction. We note, however, that this needs to be done in general only for the fiducial cosmology for which the covariance is to be computed.
        
        %%%%%%%%%%%%%%%%%%%%%%%%%%%%%%%%%%%%%%%%%%%%%%%%%%%%%%%%%%%%%%%%%%%%
    	\end{appendix}
        
\end{document}